\documentclass[iop]{emulateapj}

\newcommand{\be}{\begin{equation}}
\newcommand{\ee}{\end{equation}}

\begin{document}

\title{Implications of the Anomalous Outburst in the Blazar PKS 0208-512}

\author{Ritaban Chatterjee\altaffilmark{1}, Krzysztof Nalewajko\altaffilmark{2}, Adam D. Myers\altaffilmark{1}}

\altaffiltext{1}{Department of Physics and Astronomy 3905, University of Wyoming, 1000 East University, Laramie, WY 82071; rchatter@uwyo.edu}
\altaffiltext{2}{University of Colorado, 440 UCB, Boulder, CO 80309, USA}

\begin{abstract}
The flat spectrum radio quasar (FSRQ) PKS~0208-512 underwent three outbursts at the optical-near-infrared (OIR) wavelengths during 2008-2011. The second OIR outburst did not have a $\gamma$-ray counterpart despite being comparable in brightness and temporal extent to the other two. We model the time variable spectral energy distribution of PKS~0208-512 during those three flaring episodes with leptonic models to investigate the physical mechanism that can produce this anomalous flare. We show that the redder-when-brighter spectral trend in the OIR bands can be explained by the superposition of a fixed thermal component from the accretion disk and a synchrotron component of fixed shape and variable normalization. We estimate the accretion disk luminosity at $L_{\rm d} \simeq 8\times 10^{45}\;{\rm erg\,s^{-1}}$. Using the observed variability timescale in the OIR band $t_{\rm var,obs} \simeq 2\;{\rm d}$ and the X-ray luminosity $L_{\rm X} \simeq 3.5\times 10^{45}\;{\rm erg\,s^{-1}}$, we constrain the location of the emitting region to distance scales that are broadly comparable with the dusty torus. We show that variations in the Compton dominance parameter by a factor of $\sim 4$ --- which may result in the anomalous outburst --- can be relatively easily accounted for by moderate variations in the magnetic field strength or the location of the emission region. Since such variations appear to be rare among FSRQs, we propose that most $\gamma$-ray/OIR flares in these objects are produced in jet regions where the magnetic field and external photon fields vary similarly with distance along the jet, e.g., $u_{\rm B}' \propto u_{\rm ext}' \propto r^{-2}$. 
\end{abstract}


\section{Introduction}
The observing strategy of the \emph{Fermi} Large Area Telescope (LAT) of scanning the sky every three hours --- combined with supporting, multi-wavelength monitoring by numerous research groups --- continues to provide flux and spectral variability information for a large sample of blazars in unprecedented detail \citep[e.g.,][]{nol12}. In the second LAT active galactic nuclei (AGN) catalog, $\gtrsim$95\% of all sources are confirmed or candidate blazars \citep[2LAC,][]{ack_2yrAGN}. Modeling of truly simultaneous time-variable spectral energy distributions (SEDs) from radio to $\gamma$-rays is the holy grail of blazar physics. Such analysis for a large sample of blazars is now possible with these data. In this paper, we model the optical-near-infrared (OIR) to $\gamma$-ray SED of the blazar PKS 0208-512 during 2008--2011 to investigate the physical parameters related to its emission.

PKS 0208-512 is a Flat Spectrum Radio Quasar (FSRQ) at redshift z=1.003 \citep{hea08}. It is detected at a 36-$\sigma$ level in the \textit{Fermi} 2-yr catalog \citep[2FGL;][]{nol12} and was regularly observed by the Yale/SMARTS optical/near-infrared monitoring program \citep{bon12,cha12}. \citet[][hereafter Paper I]{cha13} showed that between 2008 August and 2011 September, PKS 0208-512 underwent three OIR outbursts of at least 1.3 magnitudes and spanning 1 month or more. In contrast, at GeV energies the source shows similar flares only during intervals 1 and 3.

During the \emph{Fermi} era, the GeV and OIR variability of blazars have been shown to be well-correlated in most cases, which is consistent with the so called ``leptonic scenario.'' The OIR emission in blazar jets is believed to be due to synchrotron radiation from the relativistic electrons in the jet \citep{urr82,imp88,mar98}. In the leptonic model, the $\gamma$-rays are generated by the same electron distribution by inverse-Compton up-scattering of synchrotron photons from the jet itself \citep[``synchrotron self-Compton'' or SSC process;][]{mar92,chi02} or external photons from the accretion disk, broad emission lines, or the torus \citep[``external-Compton'' or EC process;][]{sik94,bla00,der09}. In both SSC and EC scenarios, a strong correlation between the variations in OIR and $\gamma$-ray emission is predicted since they are produced by the same electrons. Hence, the OIR flare without any corresponding GeV variability observed for PKS 0208-512 appears to be anomalous and deserves a more detailed study. In this paper, we model the OIR to $\gamma$-ray SED of the blazar PKS 0208-512 during the three OIR outbursts to identify the physical scenario causing the anomalous flare.

\section{GeV, Optical--Near--IR and X-ray Data}

\begin{figure*}
\epsscale{1.0}
\plotone{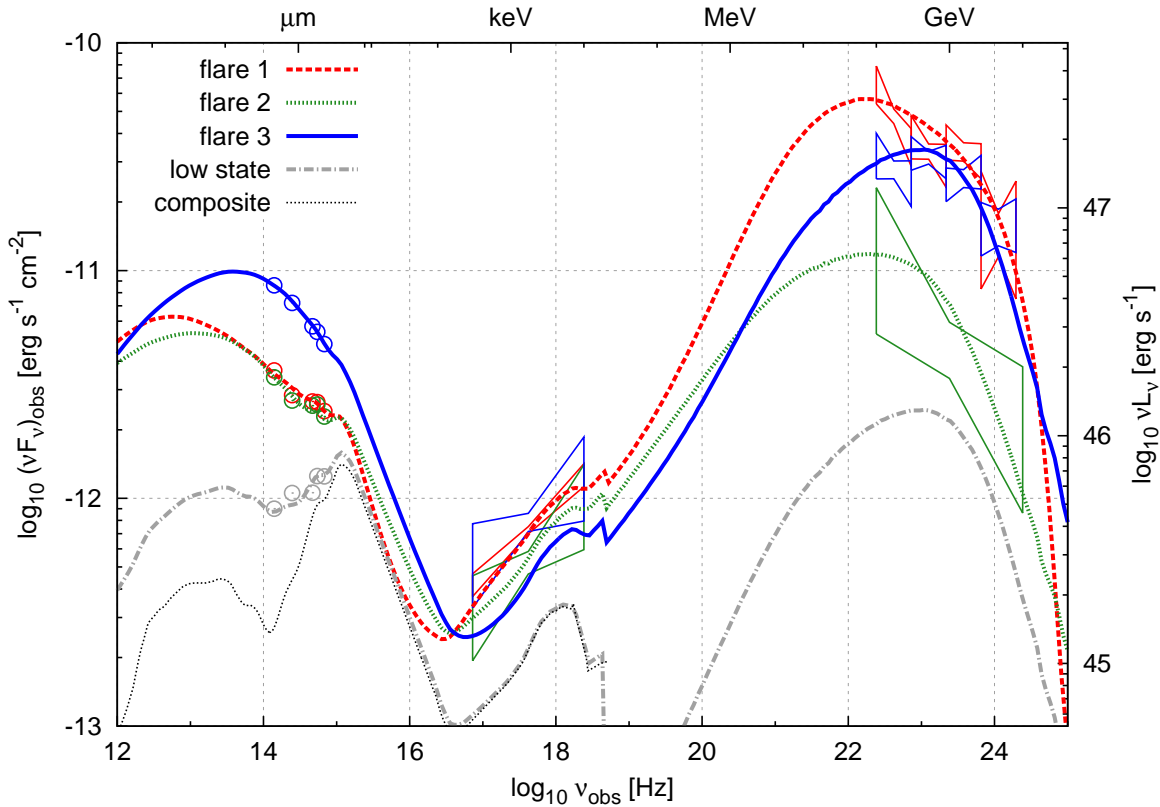}
\caption{Spectral energy distributions of PKS 0208-512, extracted from the SMARTS, Swift/XRT and Fermi/LAT data for three flaring states: \#1 (MJD 54710 -- 54840; red), \#2 (MJD 55170 -- 55230; green), \#3 (MJD 55650 -- 55850; blue); and a low state (MJD 55870 -- 55915; gray). Thick lines show leptonic models matched with the data. The black dotted line shows the quasar composite spectrum from \cite{elv94}, normalized to the low-state data.}
\label{fig_sed}
\end{figure*}

We analyzed data from \textit{Fermi}/LAT using the standard \textit{Fermi} Science Tools (v9r27p1) to derive the $\gamma$-ray spectra of PKS 0208-512. We included all sources within 15$^\circ$ of PKS 0208-512 extracted from the \textit{Fermi} 2-yr catalog (2FGL). We kept their normalizations free and spectral indices frozen to their catalog values. 
We selected the good time intervals by using the logical filter ``\texttt{DATA\_QUAL==1 \& LAT\_CONFIG==1 \& ABS(ROCK\_ANGLE) $<$ 52.}" These data were analyzed with an unbinned likelihood analysis method using the standard analysis tool \texttt{gtlike}. We use the currently recommended set of instrument response functions (\texttt{P7SOURCE\_V6}), Galactic diffuse background model (\texttt{gal\_2yearp7v6\_v0.fits}), and isotropic background model (\texttt{iso\_p7v6source.txt}). To obtain spectra, we divide the photon energy range of 0.1--20 GeV into five spectral bins and model PKS 0208-512 with a simple power law in each bin, with the spectral index and normalization kept free. The spectral bins were decided by optimizing two quantities --- namely, resolution and signal-to-noise --- in each bin. The resultant spectral data are shown in Fig.\,\ref{fig_sed}.

The goal of the SED analysis in this paper is to understand i) what physical processes could explain the differences in the $\gamma$-ray/OIR luminosity ratio between the three flares, and ii) the relation between the OIR colors and the OIR fluxes. Hence, we calculate the spectra during these three intervals. The three intervals, as defined in Paper I, included some low-state data points. This is not suitable for the current analysis since our goal is to model the SEDs during the OIR outbursts. The intervals are now defined as: 1) MJD 54710 -- 54840, 2) MJD 55170 -- 55230, and 3) MJD 55650 -- 55850, such that the low-state data points are excluded. The length of the intervals are rather long which is useful in obtaining good signal-to-noise ratios in the spectral data. However, even with such long time intervals, the source is not significantly detected (with a detection criterion that the maximum-likelihood test statistic or TS exceeds 25) in the highest energy spectral bin 8.1--20 GeV. In addition to these three intervals, we have calculated the OIR SED during a very low state MJD 55870 -- 55915. Assuming that the contribution of the variable jet flux is extremely low during this time interval, the SED is dominated by thermal emission from the accretion disk and may thus yield a relatively precise estimate of the disk emission. That estimate is crucial in setting the scale of the broad-line region (BLR) and the dusty torus, which should be the main sources of external radiation for Comptonization into $\gamma$-rays.

All measurements in the $B$, $V$, $R$, $J$, and $K$ bands were obtained with the ANDICAM instrument on the SMARTS 1.3m telescope located at CTIO, Chile \citep{bon12}. We calculate the OIR SEDs by averaging the flux over each of the three intervals including only nights when data in all 5 bands are available. 

We use the \emph{Swift/XRT data products generator} \citep{eva09} to determine the X-ray spectra in the 0.3--10~keV energy range during the three relevant intervals.

\section{SED Modeling and Results}

The selected SEDs of PKS~0208-512 are modeled with the leptonic radiative code {\tt Blazar} \citep{mod03}, which includes synchrotron emission and Comptonization of synchrotron and external radiation fields. We begin by determining the luminosity of the thermal component produced by the accretion disk. We use the composite spectrum of radio-loud quasars from \cite{elv94}, and we determine its normalization to best fit the observed low-state OIR SED. As can be seen in Fig.\,\ref{fig_sed}, and as was discussed in detail in Paper I, PKS~0208-512 shows a systematic softening of the OIR spectrum with increasing OIR flux-level, which is the redder-when-brighter behavior observed in other FSRQs \citep[e.g.][]{vil06,bon12}. We checked whether this spectral transition can be explained by the superposition of synchrotron and thermal SED components. We identified the normalization of the composite spectrum, for which the observed spectral transition is accurately reproduced by solely varying the normalization of the synchrotron component. By integrating the luminosity of the UV-peaking thermal component of the normalized composite spectrum, shown in Fig.\,\ref{fig_sed}, we find an accretion disk luminosity of $L_{\rm d} \simeq 8\times 10^{45}\;{\rm erg\,s^{-1}}$. Our value is lower by a factor of $\simeq 2.2$ than the value adopted by \cite{ghi11} which can be explained by variability of the accretion disk emission on yearly timescales.
\begin{table*}
\centering
\caption{Parameters of the SED models of three flaring states of PKS~0208-512.}
\begin{tabular}{ccccccccccc}
\hline\hline
flare & $r$ [pc] & $\Gamma_{\rm j}$ & $\xi_{\rm IR}$ & $B'$ [G] & $p_1$ & $p_2$ & $\gamma_{\rm br}$ & $\gamma_{\rm max}$ & $u_{\rm e}' [\rm erg\,cm^{-3}]$ & $N_{\rm e} [10^{54}]$ \\
\hline
1 & 2 & 20 & 0.1 & 0.17  & 1.5  & 2.55 & 850  & 12000 & 0.0145 & 1.03 \\
2 & 2 & 20 & 0.1 & 0.34  & 2.05 & 3.7  & 3000 & ---   & 0.0103 & 3.47 \\
3 & 2 & 20 & 0.1 & 0.275 & 1.6  & 4.1  & 4400 & ---   & 0.0063 & 0.52 \\
\hline\hline
\end{tabular}
\label{tab_params}
\end{table*}

We estimate the characteristic radius of the broad-line region $r_{\rm BEL}\simeq 0.091\;{\rm pc}$ and the inner radius of the dusty torus $r_{\rm IR}\simeq 2.2\;{\rm pc}$ (assuming a dust sublimation temperature of $T_{\rm IR} = 1200\;{\rm K}$) using scaling relations from \cite{sik09}. The energy densities of external radiation components depend on the assumed covering factors $\xi_{\rm BEL}$ and $\xi_{\rm IR}$, for which we try different values.

\begin{figure*}
\epsscale{1.0}
\plotone{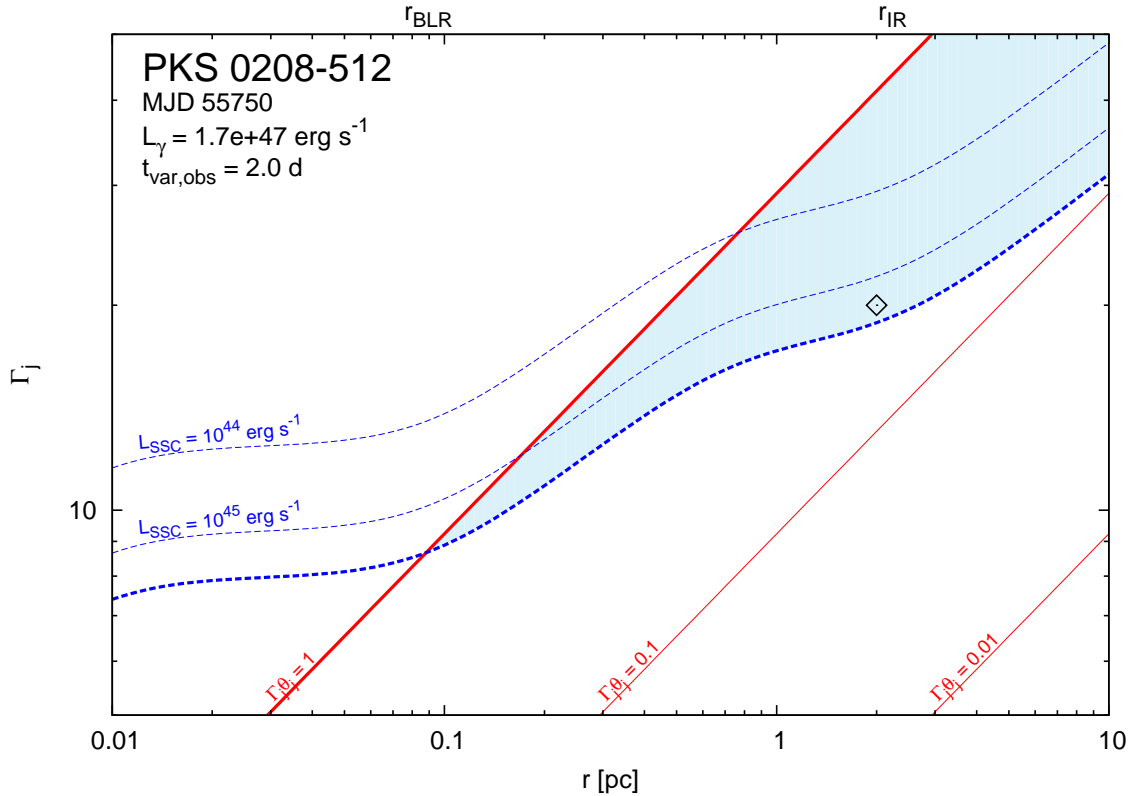}
\caption{Constraints for the location $r$ and Lorentz factor $\Gamma_{\rm j}$ of the emitting region producing the $\gamma$-ray/optical flare of PKS~0208-512 during the time interval that corresponds to flare \#3. The constraints are calculated for covering factors $\xi_{\rm BEL} = \xi_{\rm IR} = 0.1$ and a ratio of Doppler-to-Lorentz factor $\mathcal{D}/\Gamma_{\rm j} = 1.4$. Solid red lines correspond to constant $\Gamma_{\rm j}\theta_{\rm j}$, and dashed blue lines correspond to constant $L_{\rm SSC}$. The blue shaded region indicates the parameter space allowed by $\Gamma_{\rm j}\theta_{\rm j} < 1$ and $L_{\rm SSC} < L_{\rm X}$. The black diamond marks the parameters adopted in our SED modeling: $r = 2\;{\rm pc}$ and $\Gamma_{\rm j} = 20$.}
\label{fig_constraints}
\end{figure*}

We scan the OIR light curves for pairs of initial ($t_i$) and final ($t_f$) epochs such that ${\rm F_f/F_i > 2}$, where $F_i$ and $F_f$ are the corresponding fluxes, and calculate the observed variability timescale $t_{\rm var,obs}=(t_f-t_i) \times {\rm ln 2/ln(F_f/F_i)}$. We find the shortest $t_{\rm var,obs} \simeq 2\;{\rm d}$ during flares 1 and 3. We constrain the location of the emitting region $r$, and the jet Lorentz factor $\Gamma_{\rm j}=(1-\beta_{\rm j}^2)^{-1/2}$, where $\beta_{\rm j}=v_{\rm j}/c$ is the dimensionless jet velocity, corresponding to different values of the collimation efficiency parameter $\Gamma_{\rm j}\theta_{\rm j}$, where $\theta_{\rm j}$ is the jet opening angle (Nalewajko et~al., in preparation; see Appendix):
\begin{equation}
\Gamma_{\rm j}(r,\Gamma_{\rm j}\theta_{\rm j}) \simeq \left(\frac{\mathcal{D}}{\Gamma_{\rm j}}\right)^{-1/2}\left[\frac{(1+z)\Gamma_{\rm j}\theta_{\rm j} r}{ct_{\rm var,obs}}\right]^{1/2} \,,
\end{equation}
where $\mathcal{D}=[\Gamma_{\rm j}(1-\beta_{\rm j}\cos\theta_{\rm obs})]^{-1}$ is the Doppler factor, and $\theta_{\rm obs}$ is the viewing angle. There are strong theoretical and observational indications for $\Gamma_{\rm j}\theta_{\rm j} \le 1$ in AGN jets \citep{pus09,kom09}. This constraint is illustrated in Figure \ref{fig_constraints}, and one can see that it favors locations outside the BLR, although solutions located at $r_{\rm BLR}$ cannot be entirely excluded.

A further constraint can be placed using the relation between $r$ and $\Gamma_{\rm j}$ corresponding to a fixed value of the SSC luminosity $L_{\rm SSC}$ (Nalewajko et~al., in preparation; see Appendix):
\begin{eqnarray}
\Gamma_{\rm j}(r,L_{\rm SSC}) &\simeq& 1.27\left(\frac{\mathcal{D}}{\Gamma_{\rm j}}\right)^{-1}\left[\left(\frac{L_{\rm syn}}{L_{\rm SSC}}\right)\left(\frac{L_\gamma}{\zeta L_{\rm d}}\right)\right]^{1/8}
\times\nonumber\\
&&
\left[\frac{(1+z)r}{ct_{\rm var,obs}}\right]^{1/4}\,,
\end{eqnarray}
where $L_{\rm syn}$ is the synchrotron luminosity, $L_{\rm\gamma}$ is the gamma-ray luminosity, and $\zeta$ is related to the covering factors of the sources of external radiation. The observed X-ray luminosity of PKS 0208-512 is $L_{\rm X} \simeq (3.5\pm0.3)\times 10^{45}\;{\rm erg\,s^{-1}}$, but the observed X-ray spectrum is too hard to be dominated by the SSC component. It is thus required that $L_{\rm SSC} < L_{\rm X}$. This condition is illustrated in Figure \ref{fig_constraints}, and it strongly constrains the jet Lorentz factor. For the subsequent modeling, we choose a particular solution with $r = 2\;{\rm pc}$ and $\Gamma_{\rm j} = 20$, which is allowed by the constraints obtained from both flares \#1 and \#2. In many blazar studies, the Doppler-to-Lorentz factors ratio is often assumed to be $\mathcal{D}/\Gamma_{\rm j} \simeq 1$. However, in our model located at $r = 2\;{\rm pc}$, adopting $\mathcal{D}/\Gamma_{\rm j} = 1$ would require $\Gamma_{\rm j} > 20$. Higher values of the ratio are possible up to $\mathcal{D}/\Gamma_{\rm j} \lesssim 2$ for very narrow jets pointing exactly at the observer. Here, we adopt an intermediate value --- $\mathcal{D}/\Gamma_{\rm j} = 1.4$ --- which corresponds to a $\theta_{\rm obs} \simeq 1.9^\circ$, a jet opening angle of $\theta_{\rm j} \simeq 0.67^\circ$, and a collimation factor of $\Gamma_{\rm j}\theta_{\rm j} \simeq 0.235$.


Our next step is to model the SED observed during flare \#3, which shows the highest OIR flux and a strong indication of a spectral peak within the Fermi/LAT band, at $\simeq 800\;{\rm MeV}$. We inject ultra-relativistic electrons with a broken power-law spectrum $N_\gamma\propto\gamma^{-p_i}$, with spectral indices $p_1$ for $\gamma < \gamma_{\rm br}$ and $p_2$ for $\gamma > \gamma_{\rm br}$, where $\gamma_{\rm br}$ is the break Lorentz factor. The observed OIR spectrum can be used to precisely determine $p_2$, and the gamma-ray spectral peak can be used to determine $\gamma_{\rm br}$. There is no strong constraint on $p_1$, so we take the maximum value allowed by the observed X-ray emission. The observed SED places strong constraints on the Compton dominance parameter $q = L_{\rm EC}/L_{\rm syn}$, and the peak frequencies ratio $w = \nu_{\rm EC}/\nu_{\rm syn}$. It can be shown that --- when the Compton scattering takes place in the Thomson regime --- every component of external radiation has a covering factor that scales as $\xi\propto q/w^2$, independent of $B'$ or $\Gamma_{\rm j}$ \citep{sik09,nal12}. If we fix $\xi$, $L_{\rm EC}$ and $\nu_{\rm EC}$, we find that $L_{\rm syn}\propto \nu_{\rm syn}^2$, a relation which is roughly perpendicular to the observed OIR spectrum, and thus the synchrotron peak can also be robustly determined. We attempted to fit the observed SED for different values of $\xi_{\rm IR}$. We find that $\xi_{\rm IR}\lesssim 0.1$, otherwise the value of $w$ is too small as compared with observations. This value of the covering factor is typical for blazar emission models.

Finally, we model the SEDs of flares \#1 and \#2, assuming that they also are produced in the IR region. These flares have very similar average OIR spectra, but they differ by almost an order of magnitude in $\gamma$-ray luminosity. The $\gamma$-ray spectrum of flare \#1 shows hints of a complex structure, which might require two spectral bumps, one peaking below $100\;{\rm MeV}$, and another peaking at $\sim 1\;{\rm GeV}$. We attempted to model this SED using a complex electron energy distribution, but we did not obtain a solution that would be acceptable both in the $\gamma$-ray and OIR bands. We note that a two-zone model, with differences in either location, magnetic field strength and/or bulk Lorentz factor, may be required to explain the details of this SED. Instead, we present a model using a broken power-law electron distribution with a cut-off at $\gamma_{\rm max}$, that reproduces the general shape of the $\gamma$-ray spectrum. The $\gamma$-ray spectrum of flare \#2 is poorly constrained, therefore we focus on the OIR spectrum. Our best-fit models of flares 1, 2, and 3 are shown in Fig.\,\ref{fig_sed} and the corresponding parameters are presented in Table\,\ref{tab_params}.

In Table \ref{tab_params}, we also report the co-moving energy density of electrons $u_{\rm e}'$ and the total number of electrons $N_{\rm e}$. The numbers suggest that flare \#1 requires the highest $u_{\rm e}'$, and flare \#2 requires the highest $N_{\rm e}$. These numbers are, however, strongly dependent on the low-energy slope of the electron distribution $p_1$, which is the most uncertain parameter of SED models, especially for flare \#2.

\section{Discussion and Conclusions}

The main problem that we address in this work is the difference between the Compton dominance $q$ among the three flares of PKS~0208-512. In the framework of external-Compton (EC) models, we expect that $q = L_\gamma/L_{\rm IR} \simeq L_{\rm EC} / L_{\rm syn} \simeq u_{\rm ext}'/u_{\rm B}'$. Our modeling results correspond to $q_1 \simeq 20.9$, $q_2 \simeq 5.2$, and $q_3 \simeq 8.0$ for flare \#1, 2, and 3, respectively. Thus, the difference between $q$ values for flares \#1 and \#2 is a factor of $\simeq 4$. In order to explain the differences between the 3 flares of PKS~0208-512, the magnetic field strength should vary by a factor of $\simeq 2$, assuming $u_{\rm ext}'$ to be constant. Alternatively, we can assume that the magnetic field profile is fixed, with $B' \propto r^{-1}$, which means that $u_{\rm B}' \propto r^{-2}$. Since the co-moving energy density of external radiation fields is expected, in general, to have a different distance dependence, $u_{\rm ext}' \propto r^{-a'}$, we can explain these differences by varying the location of the emitting region. For example, in the case that $a' = 0$ or $a' = 4$, distance variations by a factor of $\simeq 2$ are required.

Either requirement for the differences in $q$ between the three flares of PKS~0208-512 is relatively easy to satisfy. So, it is interesting to consider how common such variations in $q$ might be between different flares produced by the same blazar, or between different blazars. Although no systematic study of the relative amplitudes of correlated $\gamma$-ray and optical flares in FSRQs has been conducted to date, our experience in working with the SMARTS and Fermi/LAT data suggests that flare \#2 is indeed anomalous. This implies that either the distance at which such flares are produced is finely tuned, or that most flares are produced in jet regions where $a' \simeq 2$. In light of the generally irregular behavior of  blazars, we are inclined to favor the latter scenario. 

The spatial distribution of the external radiation fields that are Comptonized in blazars remains poorly understood. Typically, it is assumed that the reprocessing medium is concentrated at a characteristic radius $r_{\rm ext}$ (standing for $r_{\rm BEL}$ for the BLR and $r_{\rm IR}$ for the dusty torus). Then, the distribution of external radiation in the external frame can be approximated by a broken power-law distribution, with $a \simeq 0$ for $r < r_{\rm ext}$ and $a \simeq 2$ for $r > r_{\rm ext}$. In fact, in the external frame where external radiation sources are non-relativistic and external radiation fields are not subject to extinction, we have $a \le 2$, and this limit is realized for point sources. In the co-moving frame, the distribution of external radiation can be steeper due to the Lorentz transformation $u_{\rm ext}' \simeq \mathcal{D}'^2u_{\rm ext}$, where $\mathcal{D}' = \Gamma_{\rm j}(1-\beta_{\rm j}\cos\psi)$ is a Doppler factor defined in the co-moving frame, and $\psi\simeq r_{\rm ext}/r$ for $r \gg r_{\rm ext}$ is the angle between the velocities of the external radiation photons and the emitting region in the external frame. For large $\Gamma_{\rm j}$ and small $\psi$, one can approximate $\mathcal{D}' \simeq (\Gamma_{\rm j}/2)(1/\Gamma_{\rm j}^2 + r_{\rm ext}^2/r^2)$, which highlights that there are, in fact, two regimes. For $r \ll \Gamma_{\rm j}r_{\rm ext}$, one has $\mathcal{D}' \simeq \Gamma_{\rm j}r_{\rm ext}^2/(2r^2)$, and thus $a' \simeq 6$. For $r \gg \Gamma_{\rm j}r_{\rm ext}$, one has $\mathcal{D}' \simeq 1/(2\Gamma_{\rm j})$, and thus $a' \simeq 2$. The latter regime is not very practical, because at that point the external radiation is strongly deboosted. In the former regime, one can obtain $a' < 6$ by assuming a significant spatial distribution (stratification) of the external medium emissivity $j_{\rm ext}(r)$. A distribution of external radiation with $a' \simeq 2$ can be obtained by requiring (approximately) that $j_{\rm ext} \propto r^{-2}$.

In Paper I, it was speculated that the OIR outburst during interval 2 was caused by a change in the magnetic field without any change in the Doppler factor of the emitting region or total number of emitting electrons. That will cause a synchrotron flare without any variability in the external-Compton emission. Similarly, if the magnetic field is stronger at the location of flare 2, that will result in an OIR outburst much more significant than the GeV flare and the latter may not be detected. Our analysis in this paper has demonstrated that such a scenario is a possible explanation for the anomalous flare. 

An alternative scenario was discussed in Paper I. A smaller $\Gamma_{\rm j}$ at the location of flare \#2 will result in a smaller Compton dominance causing the OIR to be more dominant than the GeV emission. The constraints on $r$ and $\Gamma_{\rm j}$, shown in Figure \ref{fig_constraints}, are inconsistent with such a scenario in which flare \#2 is produced in the inner region of the jet, where the jet is still undergoing acceleration and $\Gamma_{\rm j}$ is lower than that for the other two flares. The emitting region cannot be located at r $<$ 0.1 pc for any value of $\Gamma_{\rm j}$, unless the variability timescale is much shorter than observed (i.e. hours). However, this scenario may still be valid if the jet continues to accelerate at or beyond $\sim$ pc scale such that $\Gamma_{\rm j}$ at the location of flare 2 can be significantly smaller than those corresponding to flares 1 and 3.

\vskip 2em

This work made use of data supplied by the UK Swift Science Data Center at the University of Leicester, and of SMARTS optical/near-infrared light curves that are available at www.astro.yale.edu /smarts/glast/home.php. RC and ADM were partially supported by NASA through ADAP award NNX12AE38G and EPSCoR award NNX11AM18A. KN acknowledges support by the NSF grant AST-0907872, the NASA ATP grant NNX09AG02G, the NASA Fermi GI program, and the Polish NCN grant DEC-2011/01/B/ST9/04845.

\vskip 4em

\appendix

\section{Jet collimation constraint}

We assume that the emitting region has characteristic size $R$, which is related to the co-moving variability timescale via $R \simeq ct_{\rm var}'$. The variability timescale scales like $t'_{\rm var} = \mathcal{D}t_{\rm var,obs}/(1+z)$, where $z$ is the blazar redshift. We can also relate $R$ to the location of the emitting region via $R \simeq \theta r$, where $\theta$ is the opening angle of the emitting region. We distinguish this parameter from the jet opening angle $\theta_{\rm j}$, demanding that $\theta \le \theta_{\rm j}$. It is convenient to combine $\theta$ with the Lorentz factor, and use the collimation factor $\Gamma\theta$. We can now derive direct constraint on the source Lorentz factor as a function of $\Gamma\theta$:
\be
\Gamma(r,\Gamma\theta) \simeq \left(\frac{\mathcal{D}}{\Gamma}\right)^{-1/2}\left[\frac{(1+z)\Gamma\theta r}{ct_{\rm var,obs}}\right]^{1/2} \,.
\ee

\vskip 2em

\section{SSC constraint}

We calculate the luminosity of the SSC component using the relations $L_{\rm SSC}/L_{\rm syn} \simeq g_{\rm SSC}(u_{\rm syn}'/u_{\rm B}')$, where $g_{\rm SSC}$ is a correction factor (mainly due to spectal shape and source geometry) and $u_{\rm syn}' \simeq L_{\rm syn}'/(4\pi c R^2)$, where $L_{\rm syn}' = L_{\rm syn}/\mathcal{D}^4$. We assume that the gamma-ray emission is primarily due to EC process. The magnetic energy density is related to the external radiation energy density via Compton dominance parameter $q = L_\gamma/L_{\rm syn} = g_{\rm EC}(\mathcal{D}/\Gamma_{\rm j})^2(u_{\rm ext}'/u_{\rm B}')$, where $g_{\rm EC}$ is a correction factor (mainly due to Klein-Nishina effects). Energy density of external radiation in the jet co-moving frame is $u_{\rm ext}' \simeq \zeta\Gamma_{\rm j}^2L_{\rm d}/(3\pi cr^2)$, where $\zeta$ is a parameter related to the covering factor of the reprocessing medium (gas, dust) and $L_{\rm d}$ is the accretion disk luminosity. Putting these relations together, we obtain:
\be
r \simeq 2\Gamma_{\rm j}^4\left(\frac{\mathcal{D}}{\Gamma_{\rm j}}\right)^4
\left[\frac{1}{3}\left(\frac{g_{\rm EC}}{g_{\rm SSC}}\right)\left(\frac{L_{\rm SSC}}{L_{\rm syn}}\right)\left(\frac{\zeta L_{\rm d}}{L_\gamma}\right)\right]^{1/2}
\frac{ct_{\rm var,obs}}{(1+z)}\,,
\ee
\be
\Gamma_{\rm j}(r,L_{\rm SSC}) \simeq \left(\frac{\mathcal{D}}{\Gamma_{\rm j}}\right)^{-1}\left[3\left(\frac{g_{\rm SSC}}{g_{\rm EC}}\right)\left(\frac{L_{\rm syn}}{L_{\rm SSC}}\right)\left(\frac{L_\gamma}{\zeta L_{\rm d}}\right)\right]^{1/8}\left[\frac{(1+z)r}{2ct_{\rm var,obs}}\right]^{1/4}\,.
\ee
This constraint depends on parameters $\mathcal{D}/\Gamma_{\rm j}$, $L_{\rm syn}/L_{\rm SSC}$, and $\zeta L_{\rm d}$.

\vskip 2em

\section{Parameter $\zeta$}

Parameter $\zeta$ is related to the covering factors of the sources of external radiation. We take into account three types of external radiation: broad emission lines (BLR), hot dust radiation (IR) and direct accretion disk radiation. We use the following relation (Sikora et~al. 2009):
\be
\zeta(r,\Gamma_{\rm j}) \simeq \frac{(r/r_{\rm BLR})^2}{1+(r/r_{\rm BLR})^3}\xi_{\rm BLR} + \frac{(r/r_{\rm IR})^2}{1+(r/r_{\rm IR})^3}\xi_{\rm IR} + \frac{3}{4}\left(\frac{0.28R_{\rm g}}{r}+\frac{1}{4\Gamma_{\rm j}^4}\right)\,,
\ee
where R$_{\rm g}$ is the gravitational radius of the central black hole.

\end{document}